\documentclass[prl,twocolumn,showpacs,preprintnumbers,amsmath,amssymb,superscriptaddress]{revtex4}
\usepackage{amsfonts}

\usepackage{graphicx} 
\usepackage{dcolumn}  
\usepackage{bm}       

\newcommand{\hfbax}{\sc hfb-ax}

\newcommand{\rr} {\boldsymbol{r}}
\newcommand{\bea}{\begin{eqnarray}}
\newcommand{\eea}{\end{eqnarray}}

\begin{document}

\title{Competition between Normal Superfluidity and Larkin-Ovchinnikov
Phases of Polarized Fermi Gases in Elongated Traps}

\author{J.C. Pei}
\affiliation{Joint Institute for Heavy Ion Research, Oak Ridge
National Laboratory, Oak Ridge, TN 37831}
 \affiliation{Department of Physics and
Astronomy, University of Tennessee Knoxville, TN 37996}
\affiliation{Physics Division, Oak Ridge National Laboratory,  Oak
Ridge, TN 37831}
\author{J. Dukelsky}
\affiliation{Instituto de Estructura de la Materia, CSIC, Serrano
123, 28006 Madrid, Spain}
\author{W. Nazarewicz}
\affiliation{Department of Physics and Astronomy, University of
Tennessee Knoxville, TN 37996} \affiliation{Physics Division, Oak
Ridge National Laboratory, Oak Ridge, TN 37831}
\affiliation{Institute of Theoretical Physics, Warsaw University,
ul. Ho\.{z}a 69, PL-00681 Warsaw, Poland}

\date{\today}
\begin{abstract}
By applying the recently proposed antisymmetric superfluid local density approximation (ASLDA) to strongly interacting polarized atomic gases at unitarity in very elongated traps, we find families of Larkin-Ovchinnikov (LO) type of solutions with prominent transversal oscillation of pairing potential. These LO states coexist with a superfluid state having a smooth pairing potential. We suggest that the LO phase could be accessible experimentally  by increasing adiabatically the trap aspect ratio. We show that the local asymmetry effects contained in ASLDA do not support a deformed superfluid core predicted by previous Bogoliubov-de Gennes treatments.
\end{abstract}


\maketitle

The $s$-wave pairing in two-component fermion systems is the most
common mechanism inducing superconductivity or superfluidity in the
realm of condensed matter, nuclear physics, and cold atom
gases~\cite{Rev}. The latter system, that was experimentally
realized in recent
years~\cite{ketterle,Par06a,Shin06,Par06b,salomon}, has the great
advantage that the pairing strength can be manipulated using
Feshbach resonances to access the whole range of the BCS to BEC
crossover.  The competition between $s$-wave pairing strength, trap
deformation,  population imbalance, and temperature  could give rise
to different superconducting phases and regions of phase separation.
Recent experiments with two-component polarized ultracold atomic
gases in elongated traps open the possibility of the practical
realization of long-sought exotic superfluid phases. Among them, the
Fulde-Ferrer (FF)~\cite{FF} state representing a condensate of
Cooper pairs with a finite center-of-mass momentum and the
Larkin-Ovchinnikov (LO)~\cite{LO} state with a spatially oscillating
pairing potential have attracted much
interest~\cite{Miz05,bulgac08}.

The  Rice experiment~\cite{Par06b} on a polarized  unitary gas has
been at the center of an intense debate. They observed a spherical
distortion of the superfluid core shape with respect to the
elongated trap with an aspect ratio $\eta$$\sim$50. Such a shape
mismatch  was not observed in the MIT experiment performed with a
much less elongated trap
 ($\eta$$\sim$5) \cite{Shin06}.  Recent experiments performed in traps
  of $\eta$$\sim$22 \cite{salomon} seem to confirm the MIT results. These
  seemingly  conflicting experimental results have motivated significant
  theoretical efforts~\cite{Sen07,Ku09,baksmaty,tezuka}.

Most of the theoretical approaches to trapped cold atoms have been
based on the Local Density Approximation (LDA) or in the
Bogoliubov-de Gennes mean-field approximation (BdG) with a contact
interaction. These approximations,
 which are appropriate in the weak-coupling BCS region, are not expected to
 perform well in the strongly interacting region due to in-medium screening
 effects. However, most of the experiments are accomplished for strongly
 coupled dilute fermionic superfluids in the unitary limit in which  the
 $s$-wave scattering length approaches infinity ($|a_s|\rightarrow\infty$).
 In this limit, the system exhibits universal behavior governed by the
 densities, making it ideally suited for a
density functional theory (DFT)
description~\cite{BulYu03,Bulgac07,papen05}. DFT  offers a way to
incorporate experimental information, and the results of large-scale
calculations like Monte Carlo~\cite{Carlson03,Batr08} within an
accurate energy density able to describe strongly interacting
systems due to the incorporation of many-body effects.

 Recently, SLDA~\cite{BulYu03,Bulgac07}, a superfluid
extension of DFT, was applied to symmetric two-component systems.
Subsequently, it was generalized to polarized fermionic systems
(ASLDA)~\cite{Bulgac08b}, finding strong evidence for the existence
of a stable LO phase in the thermodynamic limit. The LO phase was
predicted to be stable  in large regions of the parameter space of a
polarized one-dimensional (1D)  system~\cite{Kaka}, and very
recently experimental evidence for a  LO phase has been obtained in
a spin mixture of ultracold $^{6}$Li atoms trapped in an array of 1D
tubes~\cite{Liao}. Extremely elongated traps offer, therefore,  a
unique opportunity to explore the transition from 3D to 1D polarized
systems, and the competition between normal and exotic superfluid
phases as a function of the aspect ratio.

In this work we study  cold polarized atoms in elongated traps at unitarity.
The self-consistent BdG equations of superfluid DFT are solved using the
coordinate-space axial solver {\hfbax}~\cite{hfbax, jcpei},  based on B-splines,
which has been demonstrated to provide precise results at  large deformations.
In a previous work~\cite{jcpei} we showed the appearance of  phase
separation in deformed  SLDA. Here, we compare SLDA results with those of the
ASLDA formalism that has been tailored to describe imbalanced systems.
Our focus is on phase separation effects, deformation of superfluid cores,
and the appearance of superfluid oscillating phases for different trap elongations.

The grand-canonical energy density functional of ASLDA can be written as~\cite{Bulgac08b}
\begin{equation}\label{EDF}
\mathcal {E}=\sum_{\sigma=\uparrow,\downarrow}
\left(\alpha_\sigma(x)\displaystyle\frac{\tau_\sigma}{2}
-\lambda_\sigma \rho_\sigma \right)
+\displaystyle\frac{(3\pi^2\rho)^{5/3}}{10\pi^2}\beta(x)
-\Delta\kappa,
\end{equation}
where the densities of spin-up $\rho_\uparrow(\rr)$ and spin-down
$\rho_\downarrow(\rr)$ atoms, the pairing densities $\kappa(\rr)$,
and pairing gaps $\Delta(\rr)$ are:
\begin{equation}\label{dens}
\begin{array}{l}
\rho_{\uparrow}(\rr)=\displaystyle\sum_i f_i|u_i(\rr)|^2,
$\hspace{5pt}$\rho_{\downarrow}(\rr)=\displaystyle\sum_i (1-f_i)|v_i(\rr)|^2\vspace{5pt}\\
\displaystyle\kappa(\rr)=\sum_{i} f_iu_i(\rr)v_i^*(\rr),
$\hspace{5pt}$\Delta(\rr) = \hspace{-2pt}-g_{eff}(\rr)\kappa(\rr),
\end{array}
\end{equation}
and $\rho=\rho_\uparrow + \rho_\downarrow$. In
Eqs.~(\ref{EDF}-\ref{dens}), $f_i=1/(1+\exp(E_i/kT))$, $u_i(\rr)$
and $v_i(\rr)$ are eigenvectors of the BdG Hamiltonian, and  $E_i$
is the corresponding eigenvalue. The BdG equations are solved
self-consistently with the chemical potentials $\lambda_\sigma$
($\sigma$=$\uparrow,\downarrow$) being  determined from the
particle-number constraints $N_\sigma$=Tr$(\rho_\sigma$)
\cite{Sen07,Ber09}, where $N_\uparrow$
($N_\downarrow$=$N-N_\uparrow$) is the number of spin-up (spin-down)
fermions. The local polarization is given by
$x(\rr)=\rho_\downarrow(\rr)/\rho_\uparrow(\rr)\leqslant 1$ while
the total  polarization of the system is
$P$=$(N_\uparrow-N_\downarrow)/N$. The quantities
$\alpha_{\sigma}(x)$ and $\beta(x)$ that parametrize the local
effective masses and normal interaction, respectively, were fitted
to Monte Carlo data and experiment~\cite{Bulgac08b}.

The minimization of the energy density (\ref{EDF}) leads to  a BdG set of equations:
\bea
  \left[
\begin{array}{clrr}%
 h_\uparrow(\rr)-\lambda_{\uparrow} & {\hspace{0.7cm} } \Delta(\rr) \\
 \Delta^*(\rr) & -h_\downarrow(\rr) +\lambda_\downarrow
\end{array}
\right]\left[
\begin{array}{clrr}%
u_i(\rr) \\
v_i(\rr)
\end{array}
\right]=E_i\left[
\begin{array}{clrr}%
u_i(\rr) \\
v_i(\rr)
\end{array}
\right]\label{BdG} \eea
where the single-particle Hamiltonian
$h_\sigma=-\frac{\hbar^2}{2m}\boldsymbol{\nabla}\cdot(\boldsymbol{\nabla}\alpha_\sigma)+U_\sigma+V_{\rm
ext}$ depends on the kinetic term, Hartree potential $U_\sigma$, and
external harmonic trap potential $V_{\rm ext}$.
The ASLDA equations are solved with a finite energy cutoff $E_c$
($|E_i|\leqslant E_c$)
and a local regularized pairing interaction $g_{eff}(\rr)=
\gamma[\rho^{1/3}(\rr)+\Lambda(\rr)\gamma]^{-1}$
\cite{Bulgac07},
where $\gamma$ is the original pairing interaction, and $\Lambda(\rr)$  is the
average regulator for the spin-up and spin-down components. The SLDA
formalism can be obtained from ASLDA  assuming $x(\rr)=1$,
resulting in identical effective masses and Hartree potentials for
spin-up and spin-down species.
The potential of the confining  trap  is $V_{\rm
ext}({\bf r})=\frac{\omega^2}{2}(r^2+z^2/\eta^2) $, where $\eta$ is the aspect
ratio  defining the trap elongation. The ASLDA equations are  solved with {\hfbax}
in a discretized rectangular axial box
($r$, $z$). We work in trap units for which $\hbar=m=\omega=1$. The energy cutoff
was assumed to be $E_c>5 \lambda_\uparrow$ (in practice, it was 35 in SLDA and 25 in ASLDA).

\begin{figure}[th]
\includegraphics[width=0.95\columnwidth]{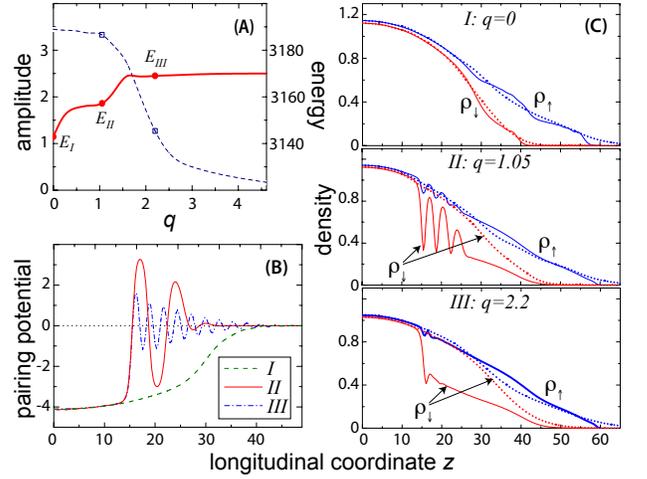}
\caption{\label{3-solution}  (Color online) SLDA results for a system of
1000 atoms in a trap with aspect ratio $\eta$=20 and
polarization of $P$=0.37.
(A) Energies of various self-consistent solutions of SLDA obtained by starting
from oscillating initial conditions with different values of $q$ (solid line).
For each solution, the amplitude of the first oscillation in the pairing potential is shown (dashed line).
(B) Pairing potentials for the three solutions with $q$=0 (I),  $q$=1.05 (II), and $q$=2.2 (III).
(C)  Density profiles $\rho_\uparrow$ and $\rho_\downarrow$ for the cases (I-III)
along  $z$  (longitudinal direction; $r=0$, solid line) and $\eta r$
(transverse direction; $z=0$, dotted line). Large deviations between longitudinal
and transverse profiles are indicative of  deformation effects.
If solid and dashed profiles are close, density distributions follow the geometry of the trap.
See text for more details.}
\end{figure}
We first study the occurrence of multiple superfluid solutions  predicted in a
 BdG treatment of large-scale trapped systems ~\cite{baksmaty}. Similar to
 Ref.~\cite{baksmaty}, we parameterize an initial guess for the pairing
potential along the $z$-axis as $\sin[q(z-z_c)]e^{-(z-z_c)/\xi}$,
where $z_c$ is the critical distance at which oscillations develop.
Figure~\ref{3-solution} displays SLDA results for $N$=1000  atoms
with polarization $P$=0.37 in a trap with  $\eta$=20. (It was
advantageous to use SLDA  in this systematic study of various
self-consistent solutions as this approach is less computationally
intensive than ASLDA.) We performed systematic calculations with
different values of the wave number $q$ finding a multitude of
self-consistent SLDA solutions. The panel A of Fig.~\ref{3-solution}
shows the energies of SLDA states as a function of $q$ together with
the amplitude of the first oscillation of the pairing potential. The
ground-state (g.s.) solution has $q$=0. In the energy curve, one can
see two plateaus, the first one with 0.4$<$$q$$<$1.1 containing
nearly degenerate solutions having large oscillation amplitudes and
the second, with $q > 1.5$, containing  nearly-degenerate solutions
exhibiting low-amplitude oscillations. From them we select three
typical solutions: (I) g.s. with $q$=0, (II) excited state with
$q$=1.05, and (III) excited state with $q$=2.2. Panel B shows the
corresponding pairing potentials along the $z$-axis and panel C
displays the density profiles in longitudinal and transverse
directions for the three states.
 Solution (II), with remarkable oscillations in the pairing potential and density profiles,
 is a LO state (see discussion below). The higher-energy state (III) does not
 show appreciable oscillations in the density profiles. Unlike the
 solution I, solution II and particularly solution III show the presence of the deformation effect
 in the minority component, similar to the one observed in the Rice experiment.

The presence of nearly degenerate solutions at very large
elongations of the trap should not be surprising. Indeed, the trap
excitation along the $z$-axis is $\Delta
E_z$=$\hbar\omega_z/2$=$1/(2\eta)$. In the example of
Fig.~\ref{3-solution},
 $\Delta E_z$=0.025, which is $\sim$0.5\% of the Fermi energy. This means that
 the trap provides a quasi-continuum background of states with smoothly increasing energy and wave number.
We note that the wave number of state (II) $q_{II}\approx q_{LO}$,
where $q_{LO}=\sqrt{2\lambda_\uparrow}-\sqrt{2\lambda_\downarrow}$
is the LO wave number. All states belonging to the plateau around
solution II have very similar values of chemical potentials, hence
practically identical values of $q_{LO}$. This suggests that the LO
can be represented by a superposition of the DFT solutions with $q$
around
 $q_{LO}$ \cite{Batr08}. Indeed, it is anticipated that the inclusion of quantum
 fluctuations beyond DFT could mix nearly degenerate solutions, thus lowering the
 energy of the LO state with respect to the ground state.

\begin{figure}[t]
\includegraphics[width=0.65\columnwidth]{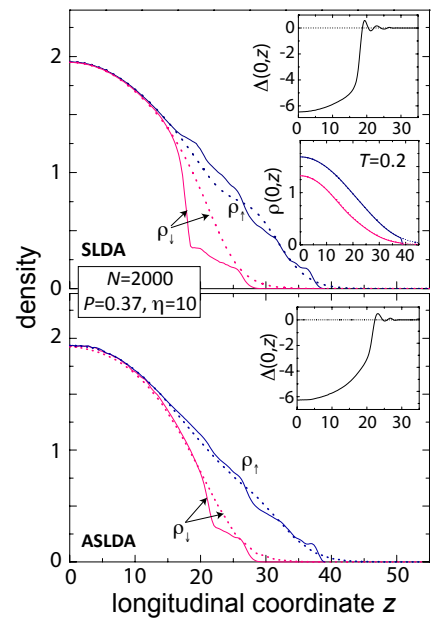}
\caption{\label{den2000} (Color online) Density profiles for the
g.s. configuration in SLDA (top) and ASLDA (bottom)  in the
longitudinal (solid lines)  and transverse (dashed lines)
directions. The pairing potentials $\Delta$ and the SLDA density
profiles at temperature $T$=0.2 are shown in the insets. }
\end{figure}

In Fig.~\ref{den2000} we compare density profiles in the longitudinal and transverse
directions in SLDA and ASLDA  for a g.s. configuration of  2000 polarized atoms with
 $P$=0.37 in a moderately elongated trap of $\eta$=10.
The insets depict the pairing potential in the longitudinal
direction. At this elongation,
 some low-amplitude oscillations  in the pairing potential are present,  but they are
 absent in the density profiles. By comparing the
longitudinal  and transverse density profiles, one can see  a clear sign of deformation
of the core in the SLDA solutions as predicted by the BdG calculations of
Refs.~\cite{Sen07,tezuka}. However, when the local polarization effects are considered
in ASLDA, the superconducting core follows the shape of the trap. Moreover, deformation
effects in SLDA are washed out at temperatures around $20\%$ of the Fermi temperature,
see Fig.~\ref{den2000}.

\begin{figure}[t]
\includegraphics[width=0.7\columnwidth]{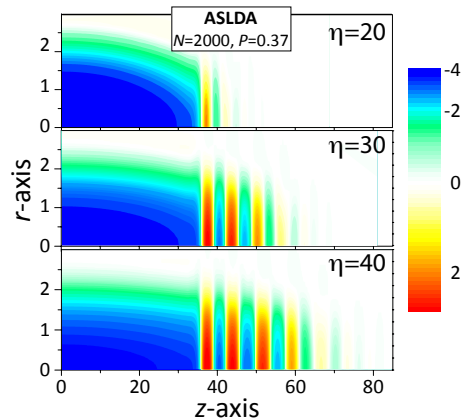}
\caption{\label{FFLO} (Color online) Pairing
potential $\Delta(r, z)$ evolved from the ASLDA g.s. solution of Fig.~\ref{den2000} to
aspect ratios $\eta$=20 (top),
30 (middle), and 40 (bottom). The radial alignment of the nodes, characteristic of
the LO state \cite{baksmaty}, is clearly seen. The LO wave numbers are $q_{LO}$=1.32,
1.18, and 1.13 for $\eta$=20, 30, and 40, respectively. Note the different scales in $z$ and $r$.}
\end{figure}

Starting from the g.s. ASLDA solution at $\eta$=10, we evolve it
gradually by increasing the aspect ratio. In this way we can reach
an excited state of the system with $q$$>$0. The resulting
two-dimensional pairing potentials  for $\eta$=20, 30, and 40 are
shown in Fig.~\ref{FFLO}. Figure~\ref{FFLO} shows that the
oscillations in the pairing potential, already present for
$\eta$=20, grow dramatically with the aspect ratio. The fact that
the oscillations have transversal character (are radially aligned),
also seen in large scale BdG calculations~\cite{baksmaty}, gives
another support for identifying this solution as a LO state.
Moreover, the strengthening of the oscillations at higher aspect
ratios is consistent with the expectation of a LO phase in the
quasi-1D limit~\cite{bulgac08,Kaka,Liao}. The numerical procedure
for the solution of ASLDA may simulate an experimental process in
which a polarized atomic gas is formed in a moderately  elongated
trap and then the aspect ratio is increased adiabatically to high
elongations.

\begin{figure}[t]
\includegraphics[width=0.9\columnwidth,height=0.5\columnwidth]{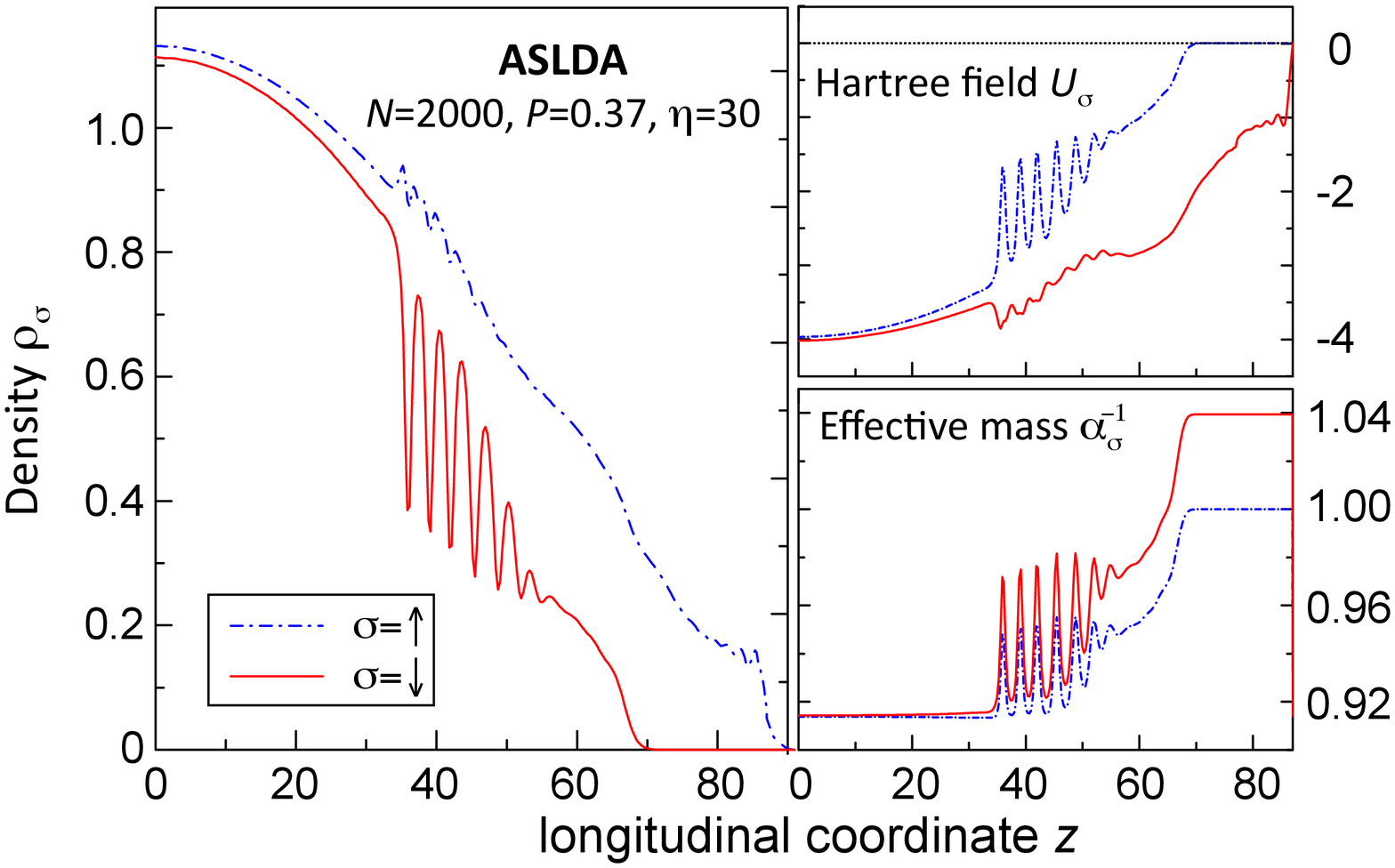}
\caption{\label{hartree} (Color online) Densities, Hartree potentials,
and effective masses for an excited ASLDA state of  2000 atoms with
$P$=0.37 in a trap with $\eta$=30. The solid and dashed-dotted lines
indicate the spin-up and spin-down components, respectively.}
\end{figure}

In order to gain a deeper insight into the ASLDA formalism, we show in
Fig.~\ref{hartree} density profiles, Hartree potentials $U_\sigma$,
and effective masses $\alpha_{\sigma}^{-1}$ in the longitudinal direction
 for the two atomic species in a trap of aspect ratio $\eta$=30 and $P$=0.37.
  All these quantities  exhibit pronounced oscillations consistent with the
  oscillations in the pairing potential of Fig.~\ref{FFLO}. Unlike in the
   SLDA formalism, these oscillations are very different for the two species.
   The Hartree potential, usually neglected in BdG calculations, has a
   significant influence on the self-consistent ASLDA solution.
   The rather shallow Hartree potential of the spin-down component does
not favor the deformed-core solution, explaining the different
density shapes seen in SLDA and ASLDA in Fig.~\ref{den2000}. In
these calculations, we used a large box with $z_{\rm max}$=90. We
checked, however, that our results for profiles inside the box, are
numerically stable with respect to small variations in  $z_{\rm
max}$. 
 Another
interesting feature is that the effective masses, fitted to Monte
Carlo calculations, are different at the trap boundaries. A recent
experiment \cite{salomon} estimated the effective mass of the Fermi
polaron to be 1.17(10), while it was fitted to be 1.04 in
Ref.~\cite{Bulgac08b}. Our calculations with the new effective mass
do not change appreciably from the presented ASLDA results.

In summary, we have performed precise superfluid-DFT studies of
spin-imbalanced Fermi gases in elongated traps at unitarity.
Calculations using  initial conditions with different pairing
oscillations yield families of coexisting pairing phases: the LDA
solution characterized by a smooth pairing potential and LO type of
solutions characterized by transversal oscillation of pairing
potential in the area of phase separation. In a theory beyond BdG,
such as the multireference DFT, these LO states can be
viewed as superpositions of DFT solutions with different $q$-values.
The deformed-core solutions predicted in SLDA  for g.s.
configurations of polarized systems  in moderately elongated traps
are absent in  ASLDA. Consequently, the presence (or absence) of
such states in experiment might provide clues about the effective
interaction in strongly interacting polarized atomic gases. Our
results suggest that LO could be studied experimentally by
elongating the traps adiabatically, starting from a moderately
elongated equilibrium state.

Discussions with A. Bulgac and M.M. Forbes are gratefully
acknowledged. This work was supported by the Office of
Nuclear Physics, U.S. Department of Energy under Contract Nos.
DE-FG02-96ER40963 and DE-FC02-07ER41457, and by grant FIS2009-07277 from the
Spanish Ministry of Science and Innovation. Computational resources
were provided by the National Institute for Computational Sciences.


\begin{thebibliography}{10}

\bibitem{Rev} R. Casalbuoni and G. Nardulli, Rev. Mod. Phys. {\bf 76}, 263 (2004);
S. Giorgini, L.P. Pitaevskii, and S. Stringari, Rev. Mod. Phys. {\bf 80}, 1215 (2008);
D.J. Dean and M. Hjorth-Jensen,  Rev. Mod. Phys. {\bf 75}, 607 (2003).

\bibitem{ketterle} W. Ketterle and M.W. Zwierlein,  Riv. Nuovo Cimento. {\bf 31}, 247 (2008).

\bibitem{Par06a} G.B. Partridge {\it et al.}, Science {\bf 311}, 503 (2006).

\bibitem{Shin06} Y. Shin {\it et al.}, Phys. Rev. Lett. {\bf 97},
030401 (2006).

\bibitem{Par06b} G.B. Partridge {\it et al.}, Phys. Rev. Lett. {\bf 97}, 190407 (2006).

\bibitem{salomon} S. Nascimbene {\it et al.}, Phys. Rev.
Lett. {\bf 103}, 170402 (2009).

\bibitem{FF} P. Fulde and R. A. Ferrell, Phys. Rev. {\bf 135}, A550 (1964).

\bibitem{LO} A.I. Larkin and Y.N. Ovchinnikov, Sov. Phys. JETP {\bf 20}, 762
 (1965).

\bibitem{Miz05} T. Mizushima, K. Machida and M. Ichioka,
Phys. Rev. Lett. {\bf 94}, 060404 (2005); J. Kinnunen, L.M. Jensen,
and P. T\"orm\"a, Phys. Rev. Lett. {\bf 96},  110403 (2006); X.J.
Liu, H. Hu, and P.D. Drummond, Phys. Rev. A {\bf 76}, 043605 (2007);
M.R. Bakhtiari, M.J. Leskinen and P. Torma, Phys. Rev. Lett. {\bf
101}, 120404 (2008); J.M. Edge and N.R. Cooper, Phys. Rev. Lett.
{\bf 103}, 065301 (2009); R.A. Molina, J. Dukelsky, and  P.
Schmitteckert, Phys. Rev. Lett. {\bf 102}, 168901 (2009).

\bibitem{bulgac08} A. Bulgac and Michael McNeil Forbes, Phys. Rev. Lett.{\bf 101}, 215301 (2008).

\bibitem{baksmaty} L.O. Baksmaty {\it et al.},
arXiv:1003.4488 (2010).

\bibitem{Sen07} R. Sensarma {\it et al.}, arXiv:0706.1741 (2007).

\bibitem{Ku09} M. Ku, J. Braun, and A. Schwenk, Phys. Rev. Lett. {\bf 102}, 255301
(2009); T.N. De Silva and E.J. Mueller, Phys. Rev. Lett. {\bf 97},
070402 (2006); M. Haque and H.T.C. Stoof, Phys. Rev. Lett. {\bf 98},
260406 (2007); M.M. Parish and D.A. Huse, Phys. Rev. A {\bf 80},
063605 (2009).

\bibitem{tezuka} M. Tezuka, Y. Yanase, and M. Ueda, arXiv:0811.1650 (2010).

\bibitem{BulYu03} A. Bulgac and Y. Yu,  Phys. Rev. Lett. {\bf 91}, 190404 (2003).

\bibitem{Bulgac07} A. Bulgac, Phys. Rev. A {\bf 76}, 040502(R) (2007).

\bibitem{papen05} T. Papenbrock, Phys. Rev. A. {\bf 72}, 041603(R) (2005).

\bibitem{Carlson03} J. Carlson {\it et al.}, Phys. Rev. Lett. {\bf 91}, 050401 (2003).

\bibitem{Batr08} G.G. Batrouni {\it et al.}, Phys. Rev. Lett. {\bf 100}, 116405 (2008).

\bibitem{Bulgac08b} A. Bulgac and M.M. Forbes, arXiv:0808.1436 (2008).

\bibitem{Kaka} A.E. Feiguin and F. Heidrich-Meisner, Phys. Rev. B {\bf 76}, 220508(R) (2007); P. Kakashvili and C.J. Bolech, Phys. Rev. A {\bf 79}, 041603(R) (2009);
M. Tezuka and M. Ueda, arXiv:1002.1433 (2010).

\bibitem{Liao} Yean-an Liao, {\it et al.}, arXiv:0912.0092 (2009).

\bibitem{jcpei} J.C. Pei, W. Nazarewicz, and M. Stoitsov, Eur. Phys. J. A {\bf 42}, 595 (2009).

\bibitem{hfbax} J.C. Pei {\it et al.}, Phys. Rev. C {\bf 78}, 064306 (2008).

\bibitem{Ber09} G. Bertsch {\it et al.}, Phys. Rev. A {\bf 79}, 043602 (2009).

\end{thebibliography}
\end{document}